\documentclass[twocolumn,aps,prb,preprintnumbers,citeautoscript,superscriptaddress,longbibliography]{revtex4-2}

\usepackage{physics}
\usepackage{makecell}
\usepackage{graphicx}
\usepackage{amsmath}
\usepackage[hidelinks]{hyperref}
\usepackage{dsfont}
\usepackage[ddmmyyyy,hhmmss]{datetime}
\usepackage{tipa}
\usepackage{booktabs}
\begin{document}

\title{Realizing a 1D topological gauge theory in an optically dressed BEC}
\author{Anika Fr\"{o}lian}\thanks{These authors contributed equally to this work}
\affiliation{ICFO - Institut de Ciencies Fotoniques, The Barcelona Institute of Science and Technology, 08860
Castelldefels (Barcelona), Spain}
\author{Craig S. Chisholm}\thanks{These authors contributed equally to this work}
\affiliation{ICFO - Institut de Ciencies Fotoniques, The Barcelona Institute of Science and Technology, 08860
Castelldefels (Barcelona), Spain}
\author{Elettra Neri}
\affiliation{ICFO - Institut de Ciencies Fotoniques, The Barcelona Institute of Science and Technology, 08860
Castelldefels (Barcelona), Spain}
\author{Cesar R. Cabrera}\thanks{Present address: Institut f\"{u}r Laserphysik, Universit\"{a}t Hamburg, Luruper Chaussee 149, 22761 Hamburg, Germany}
\affiliation{ICFO - Institut de Ciencies Fotoniques, The Barcelona Institute of Science and Technology, 08860
Castelldefels (Barcelona), Spain}
\author{Ram\'{o}n Ramos}
\affiliation{ICFO - Institut de Ciencies Fotoniques, The Barcelona Institute of Science and Technology, 08860
Castelldefels (Barcelona), Spain}
\author{Alessio Celi}\email{Electronic address: alessio.celi@uab.cat}
\affiliation{Departament de F\'{i}sica, Universitat Aut\`{o}noma de Barcelona, E-08193 Bellaterra, Spain}
\author{Leticia Tarruell}\email{Electronic address: leticia.tarruell@icfo.eu}
\affiliation{ICFO - Institut de Ciencies Fotoniques, The Barcelona Institute of Science and Technology, 08860
Castelldefels (Barcelona), Spain}
\affiliation{ICREA, Pg. Llu\'{i}s Companys 23, 08010 Barcelona, Spain}

\date{\today}

\maketitle
\textbf{Topological gauge theories describe the low-energy properties of certain strongly correlated quantum systems through effective weakly interacting models \cite{Fradkin2013, Wen2004}. A prime example is the Chern-Simons theory of fractional quantum Hall states, where anyonic excitations emerge from the coupling between weakly interacting matter particles and a density-dependent gauge field \cite{Ezawa2008}. While in traditional solid-state platforms such gauge theories are only convenient theoretical constructions, engineered quantum systems enable their direct implementation and provide a fertile playground to investigate their phenomenology without the need for strong interactions \cite{Weeks2007}. Here, we report the quantum simulation of a topological gauge theory by realising a one-dimensional reduction of the Chern-Simons theory (the chiral BF theory \cite{Rabello1995, Rabello1996, Aglietti1996}) in a Bose-Einstein condensate. Using the local conservation laws of the theory we eliminate the gauge degrees of freedom in favour of chiral matter interactions \cite{Jackiw1997, Griguolo1998, Edmonds2013, Chisholm2022}, which we engineer by synthesising optically dressed atomic states with momentum-dependent scattering properties. This allows us to reveal the key properties of the chiral BF theory: the formation of chiral solitons and the emergence of an electric field generated by the system itself. Our results expand the scope of quantum simulation to topological gauge theories and pave the way towards implementing analogous gauge theories in higher dimensions \cite{Valenti2020}.}

Gauge theories play a fundamental role in our understanding of nature, from the interactions between elementary particles to the effective description of strongly correlated systems. In recent years, there has been an immense effort to implement them in engineered quantum systems, in a quantum simulation approach \cite{Wiese2013, Zohar2016, Dalmonte2016, Banuls2020, Klco2022}. The motivation for this research is two-fold. On one hand, quantum simulators are expected to shed new light on the equilibrium and real-time dynamics of strongly-coupled gauge theories which are currently out of reach of classical computing methods, with the long-term goal of reproducing the phase diagram of quantum chromodynamics or heavy-ion collisions \cite{Fukushima2011, Berges2021}. On the other hand, they facilitate direct engineering of effective field theories that describe exotic excitations emerging in condensed-matter systems without the need for strong correlations \cite{Weeks2007}. Harnessing anyons or Majorana fermions in highly controllable, weakly interacting systems is key for investigating their potential applications, one example being topological quantum computing \cite{Nayak2008}. Significant experimental progress has been reported in both directions. Recent experiments have successfully implemented lattice gauge theories akin to the Schwinger model (1+1-dimensional quantum electrodynamics) with trapped ions \cite{Martinez2016, Kokail2019}, Rydberg atom arrays \cite{Bernien2017, Surace2020} and ultracold atoms \cite{Yang2020, Zhou2021}, and demonstrated important building blocks \cite{Dai2017, Klco2018, Gorg2019, Schweizer2019, Mil2020} to extend these schemes to higher dimensions and more complex theories. First steps have also been undertaken to realise gauge theories for anyons, demonstrating some of their key ingredients in both few-particle \cite{Roushan2017, Lienhard2020} and extended systems \cite{Clark2018, Yao2022}, although the local conservation laws required to implement a gauge invariant theory were not enforced in these experiments.

Here, we report on the quantum simulation of the chiral BF theory, a topological gauge theory introduced in the '90s to describe one-dimensional anyons in the continuum \cite{Rabello1995, Rabello1996, Aglietti1996, Kundu1999}. We experimentally investigate its peculiar phenomenology, which encompasses chiral soliton excitations and an electric field generated by the system itself, in a weakly interacting Bose-Einstein condensate (BEC) using an encoding that ensures gauge invariance by construction.

\begin{figure}[b]
\includegraphics{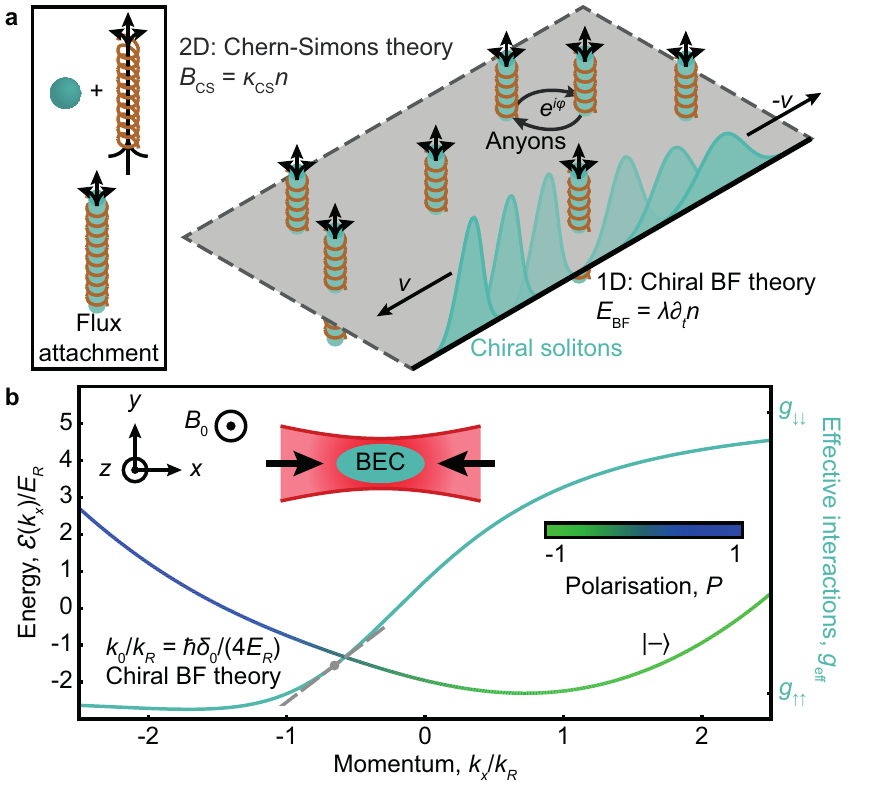}
\caption{\textbf{Simulation of the chiral BF theory in an optically dressed BEC.} \textbf{a}, The 1D chiral BF theory is a dimensional reduction of the 2D Chern-Simons theory of fractional quantum Hall states and describes chiral solitons: matter wavepackets that propagate without dispersion for only one sign of the velocity $v$ and correspond to the many-body edge states of the original 2D system. Both are topological gauge theories with local conservation laws linking the electric $E_{\mathrm{BF}}$ and magnetic $B_{\text{CS}}$ fields to $\partial_t n$ and $n$, respectively, where $n$ is the matter density. The magnetic field relation corresponds to the flux attachment condition (the formation of composite particles consisting of particles bound to magnetic flux tubes) which these theories are built upon (box) \cite{Wilczek1982}. \textbf{b}, Implementation of the chiral BF theory in the lower dressed state $\ket{-}$ of an optically coupled BEC, where states $\ket{\uparrow}$ and $\ket{\downarrow}$ are coupled by two counter-propagating Raman beams (inset). Its single-particle dispersion relation (left axis) depends on the two-photon Raman Rabi frequency $\Omega$ and detuning $\delta_0$, and its effective interaction strength $g_{\mathrm{eff}}$ (cyan, right axis) is controlled by the spin polarisation $P$ and the bare interaction parameters $g_{\uparrow\uparrow} \ll g_{\downarrow\downarrow}$. Around $k_0/k_R=\hbar\delta_0/(4E_R)$ the BEC has chiral interactions $g_{\mathrm{eff}}\propto k_x$ (grey dashed line) and realises the chiral BF theory. Here $\hbar\Omega/E_R=5.3$, $\hbar\delta_0/E_R=-2.62$, and $g_{\mathrm{eff}}$ has been computed for $^{39}$K at a magnetic field $B_0 = 374.29$~G.}
\label{Fig:Fig1}
\end{figure}

The chiral BF theory is a one-dimensional reduction of the $U(1)$ Chern-Simons theory of conventional fractional quantum Hall states \cite{Ezawa2008}. It describes a non-relativistic bosonic matter field $\hat{\Psi}$ coupled to two fields: one scalar field $\hat{\mathcal{B}}$ and one gauge field $\hat{A}$ \cite{Aglietti1996, Jackiw1997, Griguolo1998}. The gauge field is topological, i.e. it does not have propagating degrees of freedom in vacuum. It acquires dynamics from its coupling to matter through the local conservation law of the theory $\partial_x\mathcal{\hat{B}}=\kappa\hat{n}$, where $\hat{n}=\hat{\Psi}^{\dagger}\hat{\Psi}$ is the matter density and $\kappa$ is the dimensionless Chern-Simons level \cite{Chisholm2022}.  This relation is the analogue of the Gauss law in electrodynamics and defines the physical states of the theory. It translates into a relation between the expectation values of the electric field conjugated to $\hat{A}$ and the time derivative of the matter density $\langle\hat{E}\rangle=-\langle \partial_t \hat{A} \rangle=\lambda \langle\partial_t\hat{n}\rangle$, where $\lambda$ is a proportionality factor \cite{Griguolo1998, Chisholm2022}.
The chiral BF theory is a minimal model of a topological gauge theory. One of its key properties is the existence of chiral soliton solutions for the matter field: collective excitations propagating without dispersion only along one direction and corresponding to the many-body chiral edge states of the original two-dimensional (Chern-Simons) system \cite{Aglietti1996, Jackiw1997, Griguolo1998}, see Fig. \ref{Fig:Fig1}\textbf{a}.

To simulate the chiral BF theory we make use of its Hamiltonian formulation and impose the local conservation law by encoding the gauge degrees of freedom in the form of non-trivial interactions between new matter fields $\hat{\phi}=\hat{U}\hat{\Psi}$, where $\hat{U}$ is a unitary transformation (see Methods). This strategy, which has been successfully exploited to simulate the Schwinger model with trapped ions, ensures gauge invariance by construction and makes an optimal use of system resources, reducing the implementation of the gauge theory to the engineering of the corresponding interaction term \cite{Martinez2016, Muschik2017}. For the Schwinger model the encoding yields long-range (Coulomb) interactions, whereas for the chiral BF theory the resulting interactions are of finite range and are chiral. The corresponding interaction energy density reads $\mathcal{\hat{H}}_{\mathrm{int}}^{\mathrm{BF}}=\lambda \hat{\phi}^{\dagger}\hat{j}\hat{\phi}/2$ \cite{Aglietti1996, Jackiw1997, Griguolo1998}, where $\hat{j}=\hbar[\hat{\phi}^{\dagger}\partial_x\hat{\phi}-(\partial_x\hat{\phi}^{\dagger})\hat{\phi}]/(2im)$ is the normal-ordered current operator, $\hbar=h/(2\pi)$ is the reduced Planck constant, and $m$ is the mass of the matter field. This interaction term explicitly breaks Galilean invariance, as can be seen by considering a semiclassical matter wavepacket of centre of mass momentum $k$, for which $j=\hbar k n/m$ and $\mathcal{H}_{\mathrm{int}}^{\mathrm{BF}}= \lambda \hbar k n^2/(2m)$, where $n=\langle \hat{n}\rangle$ is the matter density. Thus, realising the chiral BF theory corresponds to engineering matter fields with contact interactions and a coupling constant that depends linearly on the centre of mass momentum.

This situation can be implemented in a weakly interacting BEC where two internal atomic states $\ket{\uparrow}$ and $\ket{\downarrow}$ of unequal interaction strengths $g_{\uparrow\uparrow} \neq g_{\downarrow\downarrow}$ are coupled through an external electromagnetic field with Rabi frequency $\Omega$ and detuning $\delta_0$. The resulting atom-photon dressed states $\ket{-}$ and $\ket{+}$ have modified scattering properties which depend on their spin composition, a situation which was recently investigated for radio-frequency dressing \cite{Sanz2022}. If the coupling is performed optically, e.g. with two lasers in Raman configuration counter-propagating along the $x$ axis, see inset of Fig. \ref{Fig:Fig1}\textbf{b}, a momentum $2k_R$ is transferred to the atoms along $k_x$. Here $k_R$ is the recoil momentum of a single Raman laser beam of recoil energy $E_R=\hbar^2k_R^2/(2m_0)$, where $m_0$ is the mass of the atoms. In this case, the detuning and spin composition become momentum dependent and can be described by the generalised detuning $\hbar\tilde{\delta}/E_R=\hbar\delta_0/E_R-4k_x/k_R$ and spin polarisation parameter $P=\tilde{\delta}/\tilde{\Omega}$, with $\tilde{\Omega}=\sqrt{\Omega^2+\tilde{\delta}^2}$. As a result, the dressed states acquire a momentum-dependent effective interaction strength. For $\hbar\Omega/E_R>4$, the dispersion relation of the lower dressed state $\ket{-}$ has a single minimum and is separated from the higher dressed state $\ket{+}$ by a gap $\hbar\tilde{\Omega}$ larger than all the other energy scales of the system. We thus restrict our description to state $\ket{-}$. Its effective interaction strength is given by $g_{\text{eff}}(P)=[g_{\uparrow\uparrow}(1+P)^2+g_{\downarrow\downarrow}(1-P)^2+2g_{\uparrow\downarrow}(1-P^2)]/4$, where $g_{\uparrow\downarrow}$ is the interspin interaction strength, and becomes locally linear in $k_x$, see Fig. \ref{Fig:Fig1}\textbf{b}.

\begin{figure*}[ht!]
\includegraphics{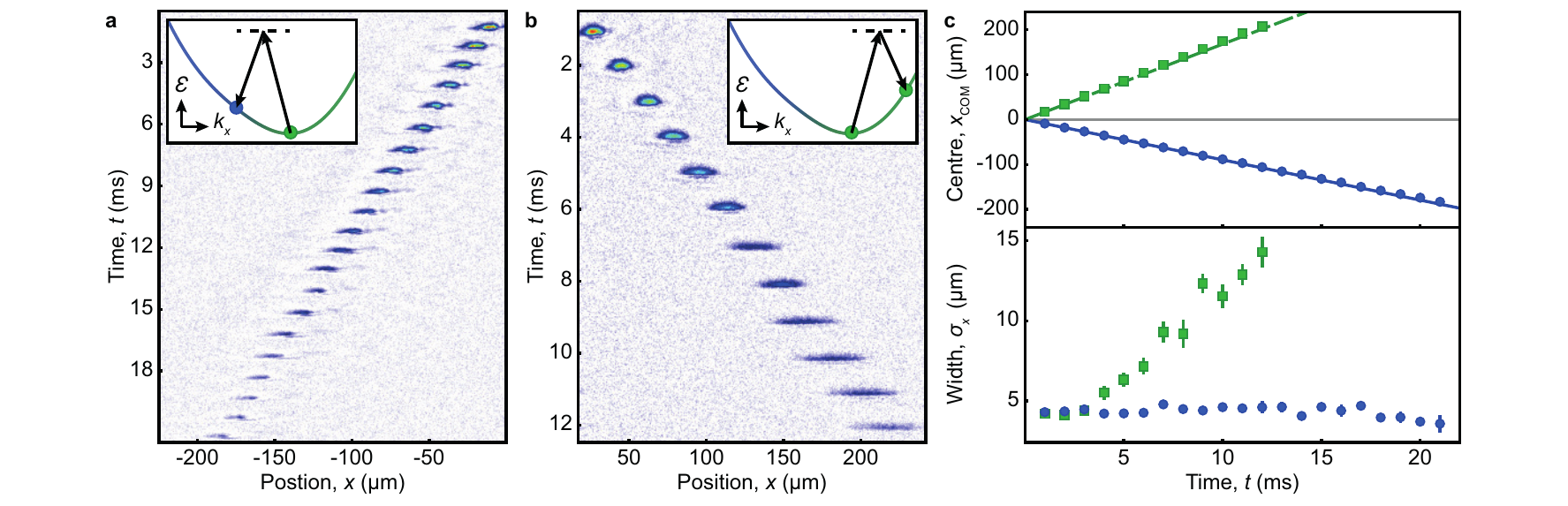}
\caption{\textbf{Observation of chiral interactions.} \textbf{a, b}, Integrated density profiles of a Raman-dressed BEC with $\hbar\Omega/E_R=5.3(3),\,\hbar\delta_0/E_R=-2.62(6),\,a_{\uparrow\uparrow}/a_0=-4.9,\,a_{\downarrow\downarrow}/a_0=24.6$, and $a_{\uparrow\downarrow}/a_0=-13.8$ measured for an evolution time $t$ in the optical waveguide. A centre of mass momentum $\Delta k/k_R=-1.45$  (\textbf{a}) or $+1.45$ (\textbf{b}) is imparted along the Raman-coupled $x$ axis using two additional Bragg beams (insets). The images correspond to single-shot measurements. \textbf{c}, Top panel: centre of mass position $x_{\text{COM}}$ of the atomic cloud vs. propagation time for $k_x>0$ (green squares) and $k_x<0$ (blue circles). The different speeds reflect the non-parabolic shape of the dispersion relation, in agreement with single-particle theory without adjustable parameters (solid and dashed lines). Bottom panel: measured cloud widths $\sigma_x$ along the waveguide direction. While the cloud expands when propagating towards the right (effective scattering length $a_{\mathrm{eff}}/a_0=21.1$), its size remains unchanged and a bright soliton forms when moving towards the left ($a_{\mathrm{eff}}/a_0=-2.7$). Values and error bars are the mean and standard deviation of three to five measurements.}
	\label{Fig:Fig2}
\end{figure*}

We exploit this linear momentum dependence to map the effective Hamiltonian of the system truncated to the lower Raman-dressed state into the quantum version of the encoded chiral BF theory \cite{Chisholm2022}. To this end, we write the effective Hamiltonian for a BEC in state $\ket{-}$ in momentum space \cite{Williams2012}, and expand it in the small parameter $(q /k_R)/[\hbar\tilde{\Omega}(k_0)/E_R]=[(k_x-k_0)/k_R]/[\hbar\tilde{\Omega}(k_0)/E_R]$, where the momentum $k_0$ must be chosen close to the centre of mass momentum of the BEC. For $\delta_0=0$ and $k_0=0$, we find
\begin{equation}
\hat{H}_{\text{eff}}\approx\int d^3\mathbf{r} \hat{\phi}^{\dagger}\left[-\frac{\hbar^2}{2}\left(\frac{\partial_x^2}{m}+\frac{\nabla_{\perp}^2}{m_0}\right)+\frac{g_1}{2}\hat{\phi}^{\dagger}\hat{\phi}+\frac{\lambda}{2}\hat{j}\right]\hat{\phi}\label{eq:Heff}
\end{equation}
which, along the $x$ direction, corresponds to the one-dimensional chiral BF Hamiltonian after encoding. Here $\nabla_{\perp}^2 = \partial_{y}^2+\partial_{z}^2$, $g_1= g_{\mathrm{eff}}(0)$, $m=m_0[1-4E_R/(\hbar\Omega)]^{-1}$ is the effective mass of the atoms along $x$, and the strength of the chiral interaction term $\lambda= mk_R(g_{\downarrow\downarrow}-g_{\uparrow\uparrow})/(m_0\Omega)$ can be experimentally controlled by adjusting the intraspin interaction strengths. Interestingly, the mapping remains valid for $\delta_0\neq 0$ provided the expansion is performed at $k_0/k_R=\hbar\delta_0/(4E_R)$, i.e. $\tilde{\delta}=0$, although in this case a static single-particle vector potential $A_{\mathrm{s}}=-\hbar k_0m/m_0$ appears. Away from $\tilde{\delta}=0$, the effective lower dressed state description remains valid but additional momentum-dependent kinetic terms beyond the chiral BF Hamiltonian need to be included (see Methods) \cite{Chisholm2022}.

The equivalence between an optically coupled BEC with unequal interaction strengths and the encoded chiral BF theory was already established in the classical field theory limit \cite{Edmonds2013}, building on the fact that, in the weakly interacting regime, the interaction term of the chiral BF theory can be recast as a density-dependent vector potential $\hat{\mathcal{A}}=-\lambda \hat{\phi}^{\dagger}\hat{\phi}/2$ \cite{Rabello1995, Rabello1996, Aglietti1996, Jackiw1997} related to the chiral BF gauge field $\hat{A}$ \emph{via} $\mathcal{\hat{A}}=\hat{A}/2$ (see Methods) \cite{Griguolo1998, Chisholm2022}. This density-dependent vector potential emerges naturally in the semiclassical description of a Raman-dressed BEC with $g_{\uparrow\uparrow}\neq g_{\downarrow\downarrow}$ due to the density-dependent detuning introduced by the differential mean-field energy shift of the transition, and can be readily calculated for large values of the Rabi frequency using a position-space approach \cite{Goldman2014}. Similarly to what happens at the single-particle level \cite{Spielman2009}, our momentum-space treatment allows one to extend the mapping to moderate values of $\Omega$, facilitating its experimental realisation.

We implement the chiral BF theory with a $^{39}\text{K}$ BEC, in which states $\ket{\uparrow}\equiv\ket{F=1,m_F=0}$ and $\ket{\downarrow}\equiv\ket{F=1,m_F=1}$ are coupled by two Raman laser beams of wavelength $\lambda_R=2\pi/k_R=768.97$ nm counter-propagating along the $x$ axis. In this configuration, the single-beam recoil energy is $E_R/h=8.66$ kHz. We apply an external magnetic field $B_0$ along the vertical $z$ direction to control the intra- and interspin interaction strengths \emph{via} the corresponding scattering lengths $a_{\uparrow\uparrow}$, $a_{\downarrow\downarrow}$, and $a_{\uparrow\downarrow}$ using Feshbach resonances. All experiments start by preparing a BEC in the lower dressed state with $\sim7000$ to $30000$ atoms, depending on the measurement. The atoms are held  in an optical dipole trap formed by crossing a waveguide beam and a confining beam, which propagate along the $x$ and $z$ axes respectively. At time $t=0$, we remove the confining beam and let the Raman-dressed cloud evolve in the waveguide, before imaging the atoms \emph{in situ} along the $z$ axis (see Methods).

To reveal the momentum-dependent nature of the interactions in this system, in a first series of experiments we investigate the dynamics of the Raman-dressed atoms when propagating in opposite directions along the $x$ axis. To this end, we prepare a BEC in state $\ket{-}$ close to the minimum of the dispersion with $\hbar\Omega/E_R=5.3(3)$, $\hbar\delta_0/E_R=-2.62(6)$, and $B_0=374.29(1)$ G (for which $a_{\uparrow\uparrow}/a_0=-4.9$,  $a_{\downarrow\downarrow}/a_0=24.6$, and $a_{\uparrow\downarrow}/a_0=-13.8$, where $a_0$ is the Bohr radius). After removing the vertical confining beam, we impart a momentum $\Delta k/k_R = \pm1.45$ to the cloud \emph{via} Bragg diffraction using two additional laser beams counter-propagating along the $x$ axis (see insets of Fig. \ref{Fig:Fig2}\textbf{a, b}). Figures \ref{Fig:Fig2}\textbf{a, b} show that the behaviour of the Raman-dressed BEC strongly depends on its propagation direction. As depicted in the top panel of Fig. \ref{Fig:Fig2}\textbf{c}, we measure a centre of mass velocity of $17.39(4)$ mm/s for a BEC moving with $k_x>0$, which is nearly twice as large in modulus as $-8.83(2)$ mm/s observed for $k_x<0$. These values agree with the single-particle theoretical prediction $v=\partial_{k_x}\mathcal{E}/\hbar$ (solid and dashed lines), and reflect the non-parabolic form of the dispersion relation $\mathcal{E}(k_x)$ at the Rabi frequency employed here. More interestingly, the width $\sigma_x$ of the atomic cloud along the propagation direction is also markedly different in the two cases. As shown in the bottom panel, for $k_x>0$ the BEC expands to more than three times its initial size in $12$ ms, whereas for $k_x<0$ it preserves its shape and $\sigma_x$ remains constant. This difference reveals the momentum dependence of interactions in our Raman-dressed system and is compatible with the effective scattering lengths $a_{\mathrm{eff}}/a_0=21.1$ for $k_x>0$ and $-2.7$ for $k_x<0$, with $g_{\mathrm{eff}}=4\pi\hbar^2a_{\mathrm{eff}}/m_0$. We attribute the absence of expansion in the attractive case to the formation of a new type of bright soliton, different from those observed so far \cite{Strecker2002, Khaykovich2002}.

\begin{figure}
\includegraphics{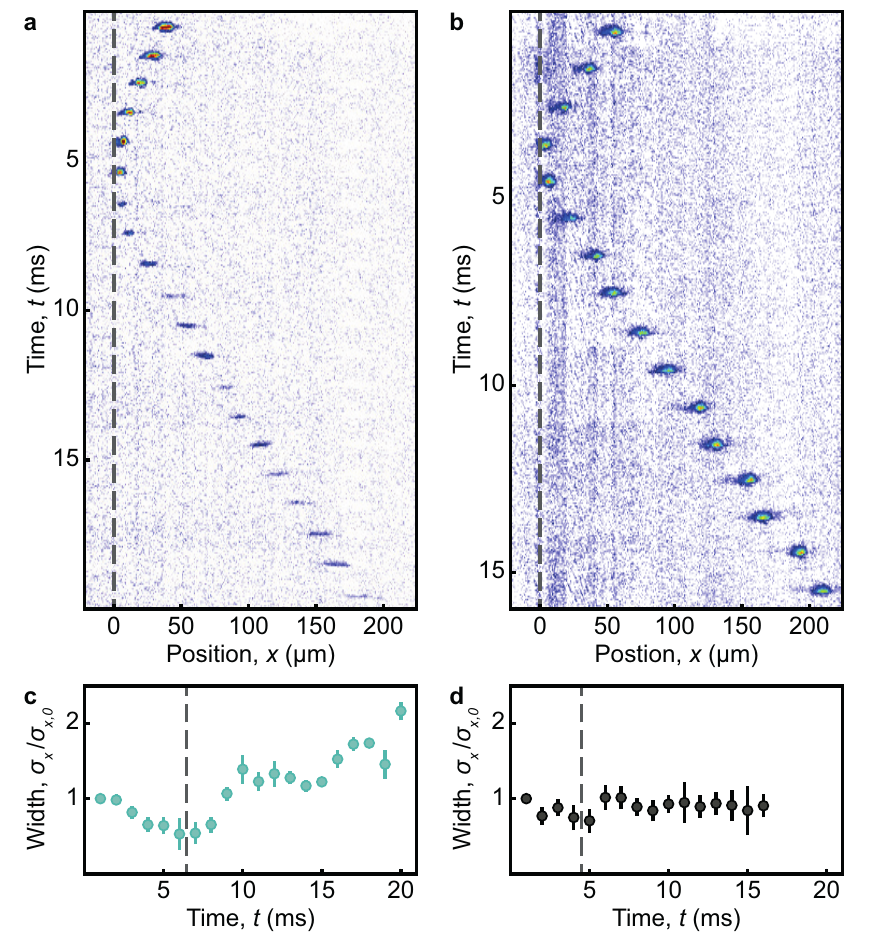}
\caption{\textbf{Chiral bright solitons.}  \textbf{a, b}, Integrated density profiles in the presence of an optical barrier (dashed line) for (\textbf{a}) a Raman-dressed soliton with the parameters of Fig. \ref{Fig:Fig2} and (\textbf{b}) a conventional bright soliton with scattering length $-2.3a_0$. Upon reflection on the barrier, the Raman-dressed soliton dissociates and starts expanding, while the conventional soliton remains unchanged. The images correspond to single-shot measurements. \textbf{c, d}, Width of the atomic cloud $\sigma_x$ renormalised by its initial value $\sigma_{x,0}$ as a function of time. Values and error bars are the mean and standard deviation of three to five measurements.}
\label{Fig:Fig3}
\end{figure}

To characterise such a Raman-dressed soliton, we investigate its behaviour after colliding with a potential barrier created by focusing a blue-detuned laser beam on the left side of the BEC’s initial position (see Methods). For barrier heights exceeding the kinetic energy of the atoms, the propagation direction of the atomic cloud is inverted by the collision \cite{Marchant2013}. As depicted in Fig. \ref{Fig:Fig3}, this process dissociates the Raman-dressed soliton making it expand (Fig. \ref{Fig:Fig3}\textbf{a, c}), while a conventional single-component bright soliton in state $\ket{\uparrow}$ with $a_{\uparrow\uparrow}/a_0=-2.3$ remains self-bound after reflection (Fig. \ref{Fig:Fig3}\textbf{b, d}). We conclude that our Raman-dressed solitons are chiral, that is, they exist only for one propagation direction. Moreover, since kinetic corrections to the encoded chiral BF theory Hamiltonian remain small for $k_x<0$ (see Methods), our Raman-dressed solitons can be considered the first experimental realisation of the chiral BF solitons theoretically predicted in refs. \cite{Aglietti1996, Jackiw1997, Griguolo1998}.

\begin{figure*}[ht]
\includegraphics{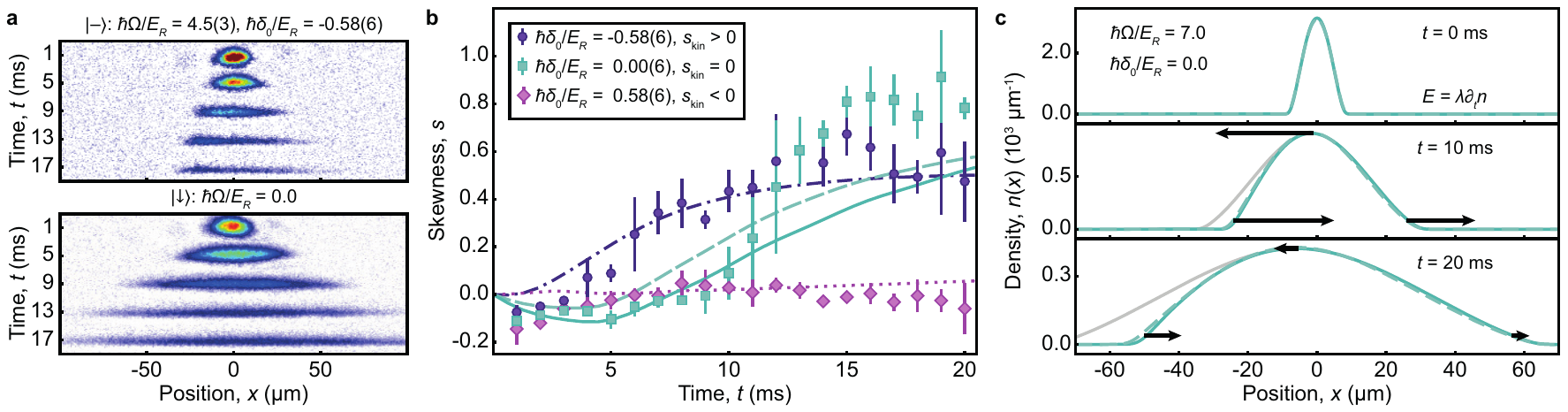}
\caption{\textbf{Revealing the chiral BF electric field.} \textbf{a}, Integrated density profiles during expansion in an optical waveguide for a Raman-dressed BEC in state $\ket{-}$ with $\hbar\Omega/E_R=4.5(3)$, $\hbar\delta_0/E_R=-0.58(6)$, $a_{\uparrow\uparrow}/a_0=1.3$, $a_{\downarrow\downarrow}/a_0=252.7$, and $a_{\uparrow\downarrow}/a_0=-6.3$ (top panel), and for a BEC in state $\ket{\downarrow}$ (bottom panel). The combination of Raman coupling and unequal interactions leads to an asymetric density profile along the $x$ direction. The images correspond to single-shot measurements. \textbf{b}, Skewness of the density profiles $s=\mu_3/\sigma^3$ for $\hbar\delta_0/E_R=\pm0.58(6)$ and $0.00(6)$, which has both kinetic $s_{\text{kin}}$ and interaction $s_{\text{int}}$ contributions linked to the non-parabolic form of the dispersion relation and to the chirality of interactions, respectively. The dashed cyan, dotted pink and dashed-dotted purple lines are numerical solutions of the two-component Raman-coupled Gross-Pitaevskii equations, and the solid cyan line corresponds to the effective chiral BF theory for $\delta_0=0$ and $k_0=0$, for which $s_{\text{kin}}=0$. Values and error bars are the mean and standard deviation of four to five measurements. \textbf{c}, Density profiles for $\hbar\Omega/E_R=7$ and $\hbar\delta_0/E_R=0$ predicted for the chiral BF theory (solid cyan line) and the two-component system (dashed cyan line), showing the validity of the mapping in this regime. The decrease in density leads to an emerging chiral BF electric field $\langle\hat{E}\rangle=\lambda\langle\partial_t\hat{n}\rangle$, which varies along the density profile (black arrows) and gives the cloud the asymmetric shape observed experimentally. Grey lines depict symmetric density profiles and serve as a guide to the eye.}
\label{Fig:Fig4}
\end{figure*}

In a second series of experiments, we investigate the other defining feature of the chiral BF theory: the emergence of an electric field generated by the system itself. Even though, by construction, the encoding eliminates the gauge degrees of freedom and encapsulates their effect into the interaction properties of the new $\hat{\phi}$ matter fields, we can access the expectation value of the chiral BF electric field through measurements of the matter field alone by using the relation $\langle\hat{E}\rangle=-\langle \partial_t \hat{A} \rangle=\lambda \langle\partial_t\hat{n}\rangle$ imposed by
the local conservation law of the theory. Experimentally, we reveal the emergence of this electric field by letting the cloud expand, which results in large temporal variations of the matter density. Concretely, we prepare a Raman-dressed system in the minimum of the dispersion relation and study its expansion dynamics in the optical waveguide focusing on the regime of repulsive effective interactions ($B_0=397.01(1)$ G, where $a_{\uparrow\uparrow}/a_0=1.3$, $a_{\downarrow\downarrow}/a_0=252.7$, and $a_{\uparrow\downarrow}/a_0=-6.3$). As shown in Fig. \ref{Fig:Fig4}\textbf{a}, for $\hbar\delta_0/E_R=-0.58(6)$ and $\hbar\Omega/E_R=4.5(3)$ the Raman-dressed BEC develops an asymmetric density profile over time. This behaviour, which does not appear in the absence of coupling, has been theoretically identified as a clear fingerprint of the chiral BF theory in the limit of large Rabi frequency \cite{Edmonds2013}.

We quantify the asymmetry of the density distribution \emph{via} the skewness parameter $s=\mu_3/\sigma^3$, where $\sigma^2$ and $\mu_3$ are the second and third central moments of the density. The evolution of $s$ during the expansion has a non-trivial dependence on detuning, which we show in Fig. \ref{Fig:Fig4}\textbf{b} for three different values of $\delta_0$. It stems from an interplay of single-particle and many-body effects that is well captured by the effective field theory of the lower dressed state but goes beyond the regime of validity of the encoded chiral BF Hamiltonian. The observed skewness has both kinetic and interaction contributions $s=s_{\text{kin}}+s_{\text{int}}$. Due to the momentum spread of the BEC, momentum-dependent terms beyond the chiral BF Hamiltonian, which can be formulated as a momentum-dependent effective mass, lead to $s_{\rm{kin}}<0$ $(>0)$ for $\hbar\delta_0/E_R>0$ ($<0$) \cite{Khamehchi2017}. Moreover, the momentum dependence of interactions, which makes the atoms expand slower when moving to the left than to the right, gives rise to $s_{\rm{int}}>0$ for any sign of $\delta_0$. Our experimental results confirm this qualitative picture, with both contributions cancelling out for $\hbar\delta_0/E_R=0.58(6)$ and adding up  when $\hbar\delta_0/E_R=-0.58(6)$. This behaviour is also well captured by Gross-Pitaevskii simulations for the Raman-coupled two-component system (lines) \cite{Chisholm2022}, which are in good agreement with the experimental measurements without adjustable parameters.

For $\hbar\delta_0/E_R=0$, the lowest order kinetic corrections to the effective chiral BF Hamiltonian of equation~\eqref{eq:Heff} cancel out, $s_{\mathrm{kin}}=0$, and our experimental observations are qualitatively reproduced by the encoded chiral BF theory. There, the discrepancy between the effective (solid line) and complete two-component (dashed line) theories stems from higher order corrections, which remain sizeable for our experimental Rabi frequency but rapidly decrease with increasing $\Omega$. As shown in Fig. \ref{Fig:Fig4}\textbf{c}, already for $\hbar\Omega/E_R=7$ the expected density profiles are essentially indistinguishable in the two cases, demonstrating the validity of our effective chiral BF theory description well beyond that of the position-space approach \cite{Edmonds2013} (see Methods). The chiral BF theory formalism provides an intuitive explanation of the experimentally observed asymmetric expansion. Upon release in the optical waveguide the atomic density decreases over time, modifying the atomic density and leading to an emerging chiral BF electric field $\langle\hat{E}\rangle=-\langle \partial_t \hat{A} \rangle=\lambda \langle\partial_t\hat{n}\rangle$. As depicted in Fig. \ref{Fig:Fig4}\textbf{c}, in an inhomogeneous system the associated electric force is spatially dependent (black arrows) and distorts the atomic density distribution during the expansion, skewing it. Therefore, the asymmetric expansion dynamics observed in the experiment reveal the chiral BF electric field.

In conclusion, we have employed an optically coupled Bose-Einstein condensate to engineer effective chiral interactions and realised the chiral BF theory, a minimal model of a topological gauge theory corresponding to a one-dimensional reduction of the U(1) Chern-Simons theory \cite{Rabello1995, Rabello1996, Aglietti1996, Jackiw1997, Griguolo1998}. Direct extensions of our work are the investigation of the chiral BF theory in an annular geometry, where the gauge degrees of freedom cannot be completely eliminated any longer and different topological configurations associated to the magnetic flux piercing the annulus emerge \cite{Edmonds2013}, or the study of its connections with the Kundu and anyon Hubbard models and their linear anyon excitations \cite{Kundu1999, Keilmann2011, Greschner2015, Straeter2016, Bonkhoff2021}. Finally, since generalising our scheme to optical fields with orbital angular momentum has been predicted to emulate flux attachment \cite{Valenti2020}, our work should be understood as the first step in the effort to simulate two-dimensional topological field theories in experimental atomic systems and investigate anyonic excitations and topological order akin to the fractional quantum Hall effect without the need for strong correlations \cite{Weeks2007}.

\vspace{3mm}
\noindent\textbf{Data availability} The datasets supporting this study are available from the corresponding authors upon request.

\clearpage

\section*{\bf METHODS}
\subsection*{Theoretical framework}
We have presented a detailed theoretical discussion on the quantum simulation of topological gauge theories, with a focus on the chiral BF theory, in ref. \cite{Chisholm2022}. We summarise the key points relevant to this work below.

\subsubsection*{Chiral BF theory}
The chiral BF theory \cite{Rabello1995, Rabello1996} is a 1+1-dimensional topological gauge theory obtained from the dimensional reduction of the 2+1-dimensional U(1) Chern-Simons theory, to which a phenomenological chiral boson term of self-dual form \cite{Floreanini1987} is added to reproduce the behaviour of the edges of the original system \cite{Aglietti1996, Jackiw1997, Griguolo1998}. In its classical version, the theory describes a non-relativistic bosonic matter field $\Psi$ coupled to two fields, the gauge field $(A^0, A^1)$ and a scalar field $\mathcal{B}$. They correspond to the $(A^0, A^1)$ and $A^2$ components of the 2+1-dimensional gauge field before dimensional reduction, respectively. Here the indices $0,1,2$ label the temporal $t$ and spatial $x$, $y$ components. To simplify the expressions, in the following we use the notation $A^1=A$.

The Lagrangian density of the chiral BF theory reads
\par\noindent\begin{multline}
\mathcal{L} =
A^0\left(\frac{\partial_x\mathcal{B}}{\kappa}- n\right)
-\frac{\mathcal{B}}{\kappa}\partial_t A + \frac{\lambda}{2 \kappa^2} (\partial_t{\mathcal{B}})(\partial_x\mathcal{B})
\\+i\Psi^*\partial_t\Psi
+\frac{1}{2m}\Psi^*(\partial_x-i A)^2\Psi
-V(n),
\label{eq:BF_2}
\end{multline}
where $\kappa$ is the dimensionless Chern-Simons level, $n=\Psi^*\Psi$ is the matter density, $\lambda$ characterises the strength of the chiral boson term, and $V(n)$ represents the matter-matter interactions,
which depend polynomially on the matter density $n$. Here and in the rest of this section, we work in units with the reduced Planck constant $\hbar$, the speed of light $c$, the vacuum permittivity $\epsilon_0$, the vacuum permeability $\mu_0$, and the electron charge $e$ set to $1$.

In this expression, $A^0$ plays the role of a Lagrange multiplier and allows us to identify the local conservation law of the theory (the analogue of the Gauss law for electrodynamics), which reads $\partial_x\mathcal{B}= \kappa n$. The equations of motion of the matter and gauge fields derived from equation \eqref{eq:BF_2} provide an additional relation between matter and gauge fields $E=\lambda \partial_t n$.

To quantise the theory, we reformulate it in Hamiltonian form. To this end, we use the Faddeev-Jackiw first-order formalism, which separates the dynamical fields from the local conservation laws of the theory by progressively eliminating the matter-dependent gauge fields at the level of the Lagrangian \cite{Faddeev1988, Jackiw1993}. This is achieved through adequate redefinitions of the matter field and results in a Hamiltonian involving only the physical degrees of freedom of the system, and where the local conservation laws are encoded in the form of non-trivial interactions between the new matter fields $\phi=U\Psi$, where $U$ is a unitary transformation \cite{Chisholm2022}.

For the chiral BF theory, we perform the redefinition of the matter field in two steps $\phi=U\Psi=U_2U_1\Psi$. The role of the first one
\par\noindent\begin{multline}
\psi = U_1\Psi= \text{exp}\left[-i\left( \int_{x_0}^{x} {\rm d}\xi \,A(\xi,t)
- \int_{t_0}^{t} {\rm d}t' A^0(x_0,t')\right.\right.
\\ \left. \left.+ \frac \lambda{2 \kappa} {\cal B}(x_0,t)\right) \right] \Psi, \label{eq:BF-gaugeT-1}
\end{multline}
where $x_0$ and $t_0$ are arbitrary reference points for the space and time coordinates, is to eliminate the gauge degrees of freedom from the Lagrangian density. However, it results in a non-local Lagrangian not easily amenable to quantum simulation and which, after canonical quantization, yields a quantum field $\hat{\psi}$ that is not bosonic. This is not surprising: the chiral BF theory was originally constructed as a model for one-dimensional (linear) anyons in the continuum \cite{Rabello1995, Rabello1996} and, after certain controversies \cite{Aglietti1996}, has been shown to correspond to a field-theoretical formulation of the Kundu linear anyon model in the regime of weak interactions \cite{Kundu1999}.

The role of the second matter field redefinition is to remove the non-locality of the Lagrangian through a Jordan-Wigner transformation
\begin{equation}
\phi = U_2\psi=\text{exp}\left[-i\frac {\lambda}2 \int_{x_0}^{x} {\rm d}\xi \,n(\xi,t)\right] \psi \label{eq:cBF-gaugeT-2},
\end{equation}
which directly yields a Lagrangian corresponding to the Legendre transform of the canonical Hamiltonian density
\begin{equation}
{\cal H} = -\frac{1}{2m} \phi^*\partial_x^2\phi+ \tilde V (n) + \frac{\lambda}2 \phi^*j\phi
\label{eq:cBF-H-enc-1},
\end{equation}
where $j=\left[\phi^*(\partial_x\phi)-(\partial_x\phi^*)\phi\right]/(2im)$ is the matter current and $\tilde{ V}(n) = V(n) + \lambda^2n^3/(8 m)$ is still polynomial in density, but includes now an additional cubic term. Equation \eqref{eq:cBF-H-enc-1} can also be rewritten as
\begin{equation}
{\cal H} = -\frac{1}{2m} \phi^*\left(\partial_x + i \frac{\lambda}2 n\right)^2\phi + V(n),
\label{eq:cBF-H-enc-An}
\end{equation}
and interpreted as a matter field $\phi$ minimally coupled to a density-dependent vector potential ${\cal A} = -\lambda n/2$ \cite{Aglietti1996, Jackiw1997, Griguolo1998}. Note however that ${\cal A}$  is not the gauge potential of the chiral BF theory before encoding $A$. Indeed, for a particular gauge choice, the classical equations of motion of the theory imply that $A=-\lambda n=2\mathcal{A}$.

The final step is to quantise equation \eqref{eq:cBF-H-enc-1}, obtaining the quantum chiral BF Hamiltonian in encoded form
\begin{equation}
\hat{H}=\int\mathrm{d}x \hat{{\cal H}} =\int \mathrm{d}x \left( -\frac{1}{2m} \hat{\phi}^{\dagger}\partial_x^2\hat{\phi}+ :\hat{\tilde V} (\hat{n}): + \frac{\lambda}2 \hat{\phi}^{\dagger}\hat{j}\hat{\phi}\right),
\label{eq:cBF-H-enc-2}
\end{equation}
where :: denotes normal ordering. This expression encapsulates the effect of the two fields $\hat{A}$ and $\hat{\mathcal{B}}$ into a non-trivial interaction term $\mathcal{\hat{H}}_{\mathrm{int}}^{\mathrm{BF}}=\lambda \hat{\phi}^{\dagger}\hat{j}\hat{\phi}/2$ between the new bosonic matter fields $\hat{\phi}$ that involves the normal-ordered current operator $\hat{j}=[\hat{\phi}^{\dagger}\partial_x\hat{\phi}-(\partial_x\hat{\phi}^{\dagger})\hat{\phi}]/(2 i m)$. Following refs. \cite{Aglietti1996, Jackiw1997, Griguolo1998}, this interaction term is named chiral because it explicitly breaks Galilean invariance. The chirality is inherited from the self-dual chiral boson term of the classical theory before encoding $(\partial_t\mathcal{B})(\partial_x\mathcal{B})$. In the quantum regime, the local conservation law of the chiral BF theory is ensured by the local symmetry generator $\hat{G}=\partial_x\hat{\mathcal{B}}-\kappa \hat{n}$, and only the eigenstates of $\hat{G}$ of eigenvalue zero are physical states. Moreover, at the quantum level the classical relation between the electric field and the matter density becomes a relation between expectation values $\langle\hat{E}\rangle=-\langle\partial_t \hat{A}\rangle=\lambda\langle\partial_t \hat{n}\rangle$. These are precisely the expressions included in the main text.

\subsubsection*{Effective chiral BF Hamiltonian description \\of a Raman-dressed BEC}
We consider a BEC subjected to two Raman beams counter-propagating along the $x$ axis (see inset of Fig. \ref{Fig:Fig1}\textbf{b}) and coupling two internal states $\ket{\uparrow}$ and $\ket{\downarrow}$ separated by a frequency difference $\Delta$. Each Raman beam imparts a recoil momentum $k_R$ along $k_x$, leaving the orthogonal directions $\mathbf{k}_{\perp}$ unaffected. We place ourselves in the rotating frame and, to simplify the expressions, we set $k_R$ and the associated recoil energy $E_R$ to 1 in this section.
\\

{\setlength{\parindent}{0pt} \textbf{Single-particle Hamiltonian.}} The single-particle Hamiltonian of the system in the basis $\ket{\sigma}\equiv\{\ket{\downarrow,k_x-1},\ket{\uparrow,k_x+1}\}$ is \cite{Spielman2009}
\begin{equation}
\hat{H}_{\mathrm{kin}} = \int\frac{\mathrm{d}^3\mathbf{k}}{(2\pi)^3}\sum_{\sigma_1,\sigma_2}\hat{\varphi}_{\sigma_1}^{\dagger}(\mathbf{k}){\cal H}_{{\rm kin}, \sigma_1,\sigma_2}\hat{\varphi}_{\sigma_2}(\mathbf{k}),
\end{equation}
where the field operator $\hat{\varphi}_{\sigma}^{\dagger}(\mathbf{k})\,(\hat{\varphi}_{\sigma}(\mathbf{k}))$ creates (destroys) a particle in state $\ket{\sigma}$ and
\begin{equation}\label{eq:Hkin-matrix}
{\cal H}_{\mathrm{kin}}=\left[\mathbf{k}^2_{\perp}+(k_x+\sigma_z)^2\right]+\frac{\Omega}{2}\sigma_x-\frac{\delta_0}{2}\sigma_z.
\end{equation}
In this expression $\sigma_x$ and $\sigma_z$ are the corresponding Pauli matrices, and $\Omega$ and $\delta_0$ are the two-photon Rabi frequency and detuning of the Raman beams.

For each value of $k_x$, ${\cal H}_{\mathrm{kin}}$ can be diagonalised by the unitary matrix
\begin{equation}
U_R =
	\begin{pmatrix}
	\sin{\theta(k_x)} &  -\cos{\theta(k_x)}\\
	\cos{\theta(k_x)} &  \sin{\theta(k_x)} \\
	\end{pmatrix},
\label{eq:U}
\end{equation}
which relates the original $\ket{\sigma}$ basis to the Raman-dressed basis $\ket{\pm}$ \emph{via} $
\hat{\phi}_{\pm}(\mathbf{k})=\sum_{\sigma} U_{R\pm,\sigma}(k_x)\hat{\varphi}_\sigma(\bf{k})\label{Eq23}$,
where $\hat{\phi}_{\pm}({\bf k})$ are the field operators in the dressed basis. The dressed basis' eigenvalues are   $E_{\pm}=\mathbf{k}_{\perp}^2+\mathcal{E}_{\pm}=\left(\mathbf{k}_{\perp}^2+k_x^2+1\right)\pm\tilde{\Omega}/2$,
with $\tilde{\Omega}=\sqrt{\Omega^2+\tilde{\delta}^2}$ and $\tilde{\delta}=\delta_0-4k_x$, and describe two energy bands of quasi-momentum ${\bf k}$. The spin composition of the dressed basis' eigenstates is momentum dependent and can be quantified by the spin polarisation parameter $P=\tilde{\delta}/\tilde{\Omega}$, since $\sin{\theta(k_x)}=\sqrt{(1-P)/2}$ and $\cos{\theta(k_x)}=\sqrt{(1+P)/2}$. For $\Omega> 4$, the lowest energy band $\ket{-}$ has a single minimum and is separated from the higher energy band $\ket{+}$ by an energy gap $\tilde{\Omega}$ larger than all other energy scales of the system, allowing the system's description to be restricted to $\ket{-}$. Thus, we drop the subscript in $\hat{\phi}_ -$.

For a BEC with centre of mass momentum $\sim k_0$, the momentum of the atoms along the Raman-coupled direction can be recast as $k_x=k_0+q$. Their single-particle energy can be expanded in the small parameter $q/\tilde{\Omega}(k_0)$ around $k_0$, yielding
\begin{equation}
{\cal H}_{\mathrm{kin}}=\mathcal{E}_-(k_0)+\left[\mathbf{k}^2_{\perp}+\frac{(q-A_{\mathrm{s}})^2}{m^*}+W\right]+\mathcal{O}\left([q/\tilde{\Omega}(k_0)]^4\right)\label{eq:Hkin},
\end{equation}
where, along the $x$ direction, the dressed atoms have the effective mass
$m^*=[1-4\Omega^2/\tilde{\Omega}(k_0)^3]^{-1}$, experience a synthetic vector potential $A_{\mathrm{s}}=-m^*[k_0+\tilde{\delta}(k_0)/\tilde{\Omega}(k_0)-8q^2\tilde{\delta}(k_0)\Omega^2/\tilde{\Omega}(k_0)^5]$,  and are subjected to a scalar potential $W=-A_{\mathrm{s}}^2/m^*$. To third order in $q/\tilde{\Omega}(k_0)$, $A_{\mathrm{s}}$ has a momentum dependence that gives a kinetic contribution to the skewness of the cloud after expansion, see Fig. \ref{Fig:Fig4}\textbf{b}. This contribution can be cancelled by setting $\tilde{\delta}=0$ ($k_0=\delta_0/4$, $P=0$).
\\

{\setlength{\parindent}{0pt} \textbf{Interaction Hamiltonian.}} The interaction Hamiltonian of the system describes two-body collisions between atoms of incoming momenta $\mathbf{k_1}$, $\mathbf{k_2}$ and outgoing momenta $\mathbf{k_3}$, $\mathbf{k_4}$, and reads
\begin{equation} \label{Hint}
\hat{H}_{\mathrm{int}} =
\int{\frac{\textrm d^3\textbf{k}_1}{(2\pi)^3}\frac{\textrm d^3\textbf{k}_2}{(2\pi)^3}\frac{\textrm d^3\textbf{k}_3}{(2\pi)^3}
\frac{\textrm d^3\textbf{k}_4}{(2\pi)^3}\hat{\cal{V}}(\mathbf{k}_1,\mathbf{k}_2,\mathbf{k}_3,\mathbf{k}_4)},
\end{equation}
with
\par\noindent\begin{multline}
\hat{\cal{V}}(\mathbf{k}_1,\mathbf{k}_2,\mathbf{k}_3,\mathbf{k}_4)=\frac{1}{2}\sum_{\sigma_1,\sigma_2}g_{\sigma_1,\sigma_2}\hat{\varphi}_{\sigma_1}^{\dagger}(\mathbf{k}_4)\hat{\varphi}_{\sigma_2}^{\dagger}(\mathbf{k}_3)\\
\hat{\varphi}_{\sigma_1}(\mathbf{k}_2)\hat{\varphi}_{\sigma_2}(\mathbf{k}_1)\delta^3(\mathbf{k}_4+\mathbf{k}_3-\mathbf{k}_2-\mathbf{k}_1)\label{eq:Vbare},
\end{multline}
where $g_{\uparrow\uparrow}$, $g_{\downarrow\downarrow}$ and $g_{\uparrow\downarrow}$ describe the interactions of the bare states.

Following ref. \cite{Williams2012}, we rewrite $\hat{\cal{V}}$ in the dressed basis and truncate it to the lower energy band
\par\noindent\begin{multline}
 \hat{\cal{V}}_{\mathrm{eff}}(\mathbf{k}_1,\mathbf{k}_2,\mathbf{k}_3,\mathbf{k}_4)=\frac{1}{2}\tilde{g}_{\mathrm{eff}}({\bf k_1},{\bf k_2},{\bf k_3},{\bf k_4})\\
 \hat{\phi}^\dagger(\mathbf{k}_4)\hat{\phi}^\dagger(\mathbf{k}_3)
 \hat{\phi}(\mathbf{k}_2)\hat{\phi}(\mathbf{k}_1)\delta^3(\mathbf{k}_4+\mathbf{k}_3-\mathbf{k}_2-\mathbf{k}_1),
\label{eq:Veff}
\end{multline}
where the effective interaction parameter is
\par\noindent\begin{multline}
	\tilde{g}_{\mathrm{eff}}(k_{1,x},k_{2,x},k_{3,x},k_{4,x})=\sum_{\sigma_1,\sigma_2}g_{\sigma_1,\sigma_2}\\
U_{R-,\sigma_1}(k_{4,x})U_{R-,\sigma_1}^\dagger(k_{2,x})U_{R-,\sigma_2}(k_{3,x})U^\dagger_{R-,\sigma_2}(k_{1,x}).
\label{eq:geff}
\end{multline}
As before, the momenta of the atoms along $x$ can be rewritten as $k_{i,x}=k_0+q_i$, and $\tilde{g}_{\mathrm{eff}}$ can be expanded in series around any momentum $k_0$. The result is
\par\noindent\begin{multline}
\tilde{g}_{\mathrm{eff}}(k_{1,x},k_{2,x},k_{3,x},k_{4,x})=\cr
\tilde{g}_{\mathrm{eff}}(k_0,k_0,k_0,k_0)+\lambda\frac{\tilde{\Omega}(k_0)}{2m^*}\sum_{i=1}^4\frac{q_i}{\tilde{\Omega}(k_0)}\\+{\cal O}\left([q_i/\tilde{\Omega}(k_0)]^2\right)\label{eq:geff}.
\end{multline}
Here the zeroth order expansion coefficient is  $\tilde{g}_{\mathrm{eff}}(k_0,k_0,k_0,k_0)=g_{\text{eff}}(P_0)$, with $P_0=P(k_0)$ the spin polarisation parameter for $k_0$, and $g_{\text{eff}}(P)=[g_{\uparrow\uparrow}(1+P)^2+g_{\downarrow\downarrow}(1-P)^2+2g_{\uparrow\downarrow}(1-P^2)]/4$ the effective interaction strength defined in the main text. The first order expansion coefficient reads
$\lambda=m^*[4\tilde{\delta}(k_0)g_{\text{eff}}(P_0)/\tilde{\Omega}(k_0) + g_{\downarrow\downarrow}(1-P_0)^2-g_{\uparrow\uparrow}(1+P_0)^2]/\tilde{\Omega}(k_0)$, which for $\delta_0=0$ and $k_0=0$ also yields the main text expression.

Finally, transforming back to position space we obtain the interaction term of the encoded chiral BF theory around $k_0$
\begin{equation}\label{eq:Hint}
    \hat{H}_{\mathrm{int}}{
    \approx \int \mathrm{d}^3\mathbf{r}\hat{\phi}^{\dagger}\left[\frac{g_{\text{eff}}(P_0)}{2}\hat{n}
		+\frac{\lambda}{2}\hat{j}\right]\hat{\phi}},
\end{equation}
where $\hat{j}=\left[\hat{\phi}^\dagger\partial_x\hat{\phi}- \left(\partial_x\hat{\phi}^\dagger\right)\hat{\phi}\right]/(i m^*)$ is the normal-ordered current operator in dimensionless units.
\\

{\setlength{\parindent}{0pt} \textbf{Effective chiral BF Hamiltonian.}} Combining the expressions of the single-particle and interaction Hamiltonian of equations \eqref{eq:Hkin} and \eqref{eq:Hint}, we obtain for $\tilde{\delta}=0$ ($k_0=\delta_0/4$, $P=0$) the effective Hamiltonian of the Raman-coupled BEC restricted to the lower energy band
\begin{equation}
\hat{H}_{\mathrm{eff}}
\approx
\int \textrm d^3 \mathbf r\hat{\phi}^\dagger\left[-\frac{\partial_x^2}{m^*}-\nabla_{\perp}^2+\frac{1}{2}\left(g_1-\frac{\lambda\delta_0}{2} \right)\hat{n}
+ \frac{\lambda}{2} \hat{j}\right]\hat{\phi},
\label{eq:Heff-full}
\end{equation}
where $g_1=g_{\text{eff}}(0)$, and we have gauged away the single-particle vector potential $A_{\mathrm{s}}$ and dropped the constant energy contributions $\mathcal{E}_-(k_0)$ and $W$. This expression is valid to third and first order in the expansion parameter $q/\tilde{\Omega}(k_0)$ for the kinetic and interaction parts of the Hamiltonian, respectively, and is identical to the encoded version of the chiral BF Hamiltonian equation \eqref{eq:cBF-H-enc-2} for $\hat{\tilde{V}}(\hat n)=(g_1-\lambda\delta_0/2)\hat{n}^2$. For $\delta_0=0$ and $k_0=0$, it yields equation \eqref{eq:Heff} of the main text.
\\

{\setlength{\parindent}{0pt} \textbf{Numerical simulations.}} The theoretical curves of Fig. \ref{Fig:Fig4} are obtained by numerically solving the three-dimensional two-component Gross-Pitaevskii equations of the Raman-coupled system and, for $\tilde{\delta}=0$, the extended Gross-Pitaevskii equation resulting from the effective chiral BF Hamiltonian of equation \eqref{eq:Heff}. We use the package XMDS2 \cite{Dennis2013} for all simulations, which include all relevant experimental parameters (trapping potential, interactions, Raman beams, initial atom number, etc.). Atom losses are not included, since we have numerically verified that the skewness of the density distribution is resilient to them \cite{Chisholm2022}. From the simulations, we conclude that for $\tilde{\delta}=0$ and our typical experimental parameters, the validity of the mapping of the Raman-coupled BEC to the chiral BF theory is excellent already at $\Omega=7$, see Fig. \ref{Fig:Fig4}\textbf{c}. In contrast, for the position-space approach of ref. \cite{Edmonds2013}, the effective mass gives a correction greater than $100~\%$  in these conditions. We present a more detailed numerical analysis of the regime of validity of our mapping for the expansion measurements in ref. \cite{Chisholm2022}.

\subsection*{Experimental methods}
\subsubsection*{Preparation and calibration procedures}

{\setlength{\parindent}{0pt}\textbf{Interactions and traps.}} We perform all experiments with a $^{39}\text{K}$ BEC in the two lowest Zeeman sublevels of the $F=1$ hyperfine manifold $\ket{\uparrow}\equiv\ket{F=1,m_F=0}$ and $\ket{\downarrow}\equiv\ket{F=1,m_F=1}$. We set the magnetic field $B_0$ in the range $374-397$ G to adjust the scattering length $a_{\downarrow\downarrow}$ using the Feshbach resonance at $402$ G. For all experiments, we use the model interaction potentials of ref. \cite{Roy2013} to predict the values of $a_{\uparrow\uparrow}$, $a_{\downarrow\downarrow}$ and $a_{\uparrow\downarrow}$. We initially confine the atoms in a harmonic trap of frequencies $(\nu_x,\nu_y,\nu_z)$ formed by two crossed beams aligned with the $x$ (waveguide beam) and $z$ (confining beam) axes. After preparation in the lower Raman-dressed state (see below), we turn off the confining beam and release the atoms into a waveguide of trapping frequencies $(\nu^0_x,\nu^0_y,\nu^0_z)$ Hz, where the remaining trapping potential along the $x$ axis $\nu^0_x=4(1)$ Hz stems from the residual curvature of the magnetic field. The trap frequencies, magnetic field, scattering lengths, and atom number used for the different experiments are summarised in Table~\ref{tab:TableExpParam}.
\\
\begin{table}[h]
\makebox[\columnwidth][c]{
\begin{tabular} {lrrrrrr}
	\toprule
	Parameter&\phantom{aa}&\makecell[l]{ Fig. \ref{Fig:Fig2}}        &\phantom{aa}& \makecell[l]{Fig. \ref{Fig:Fig3}\textbf{b}} &\phantom{aa}& \makecell[l]{Fig. \ref{Fig:Fig4}}\\
    &\phantom{aa}&\makecell[l]{ Fig. \ref{Fig:Fig3}\textbf{a}} &\phantom{aa}&                             &\phantom{aa}&                    \\
	\bottomrule
	\makecell[l]{$\nu_x$ (Hz)}&\phantom{aa}& $76(2)$     &\phantom{aa}& $76(2)$                        &\phantom{aa}&$70(1)$\\
    \makecell[l]{$\nu^0_x$ (Hz)}&\phantom{aa}&            $4(1)$      &\phantom{aa}& $4(1)$                         &\phantom{aa}&$4(1)$\\
	\hline
	\makecell[l]{$\nu_y$ (Hz)} &\phantom{aa}& $128(2)$   &\phantom{aa}& $128(2)$                       &\phantom{aa}& $147(2)$\\
	\makecell[l]{$\nu^0_y$ (Hz)}             &\phantom{aa}& $98(2)$    &\phantom{aa}& $98(2)$                        &\phantom{aa}& $129(2)$\\
	\hline
    \makecell[l]{$\nu_z=\nu^0_z$ (Hz)}  &\phantom{aa}& $51(5)$     &\phantom{aa}& $51(5)$                      &\phantom{aa}& $99(1)$\\
    \hline
	\makecell[l]{$B_0$ (G)} &\phantom{aa}& $374.29(1)$     &\phantom{aa}& $385.62(1)$                    &\phantom{aa}& $397.01(1)$\\
	\hline	
     \makecell[l]{$a_{\uparrow\uparrow}/a_{0}$} &\phantom{aa}& $-4.9$ &\phantom{aa}& $-2.3$        &\phantom{aa}& $1.3$\\
	\hline	
    \makecell[l]{$a_{\downarrow\downarrow}/a_{0}$} &\phantom{aa}& $24.6$ &\phantom{aa}& $61.0$     &\phantom{aa}& $252.7$\\
    \hline
    \makecell[l]{$a_{\uparrow\downarrow}/a_{0}$} &\phantom{aa}& $-13.8$ &\phantom{aa}& $-10.8$     &\phantom{aa}& $-6.3$\\
    \hline	
    \makecell[l]{$N (10^3)$} &\phantom{aa}& $14(4)$ &\phantom{aa}& $7(2)$                              &\phantom{aa}& $29(5)$\\
    \bottomrule
\end{tabular}}
\caption{{\bf Experimental parameters.} Trapping frequencies in the initial crossed optical dipole trap $\nu_{x,y,z}$ and in the optical waveguide $\nu^0_{x,y,z}$, magnetic field $B_0$, scattering lengths $a_{\uparrow\uparrow},a_{\downarrow\downarrow},a_{\uparrow\downarrow}$, and atom number $N$ after $1$ ms of evolution in the waveguide, for the different experiments presented in the main text.}
\label{tab:TableExpParam}
\end{table}

{\setlength{\parindent}{0pt}\textbf{Raman dressing.}}
We optically couple states $\ket{\uparrow}$ and $\ket{\downarrow}$ using a two-photon Raman scheme with two-photon Rabi frequency $\Omega$ and detuning $\delta_0$. The two Raman beams counter-propagate along the $x$ axis, see inset of Fig. \ref{Fig:Fig1}\textbf{b}, and have linear and orthogonal polarisations. Their wavelength is $\lambda_{R} = 768.97$ nm, corresponding to the tune-out value between the potassium D1 and D2 lines, for which scalar light shifts cancel.  The $1/\text{e}^{2}$ radii of the beams at the position of the atoms are $(w_{z},w_{y}) = (75, 81)\,\mu$m, sufficiently large to consider them as plane waves in the theoretical analysis (as we have verified numerically). While the frequency difference of the two Raman beams only differs from the energy splitting between states $\ket{\uparrow}$ and $\ket{\downarrow}$ by the two-photon detuning $\delta_0$, the neighbouring Zeeman sub-levels are off-resonant by $\geq 20$ MHz. Thus, our spin-$1/2$ theoretical analysis is sufficient to describe the system.

We calibrate the two-photon Raman detuning $\delta_{0}$ and Rabi frequency $\Omega$ using the $\ket{\uparrow}$ to $\ket{\downarrow}$ transition. For $\delta_0$, we drive a radio-frequency (rf) $\pi$-pulse in a thermal cloud, to avoid systematic errors introduced by the differential mean-field energy shifts. For $\Omega$, we drive resonant Rabi oscillations using the Raman beams. These methods have uncertainties of $\leq500$ Hz and $\leq6\,\%$ respectively. Table~\ref{tab:TableRamanParam} summarises the Raman dressing parameters used for the different experiments.

\begin{table}[h!]

	\makebox[\columnwidth][c]{
		\begin{tabular} {llrrrrr}
        \toprule
			
			Experiment &\phantom{a}& \makecell[c]{$\hbar\delta_{0,i}/E_R$} &\makecell[c]{$\hbar\delta_{0,f}/E_R$} & \makecell[c]{$\hbar\Omega_{f}/E_R$} & \makecell[c]{$\tau_{1}$ (ms)} & \makecell[c]{$\tau_{2}$ (ms)}\\
			\toprule
			Figs. \ref{Fig:Fig2} \& \ref{Fig:Fig3} &\phantom{aa}& $-2.62(6)$&$-2.62(6)$   & $5.3(3)$&$40$& $5$\\
			\hline
			Fig. \ref{Fig:Fig4} &\phantom{a}& $-0.46(6)$& $-0.58(6)$                         & $4.5(3)$ & $30$    & $5$\\		
                                &\phantom{a}&           & $0.00(6)$                          &          &         & \\	
                                &\phantom{a}&           & $0.58(6)$                          &          &         & \\
				
        \bottomrule
	\end{tabular}}
	\caption{{\bf Raman dressing parameters.} Detuning $\delta_{0}$ and Rabi frequency $\Omega$ of the Raman coupling field, and timescales $\tau_{1,2}$ of our preparation procedure. For our experimental configuration, $E_R/h=8.66$ kHz.}
	\label{tab:TableRamanParam}
\end{table}

We load the atoms into the lower Raman-dressed state $\ket{-}$ starting from a BEC in state $\ket{\downarrow}$ in a two-step procedure. In step (1), we fix an initial detuning $\delta_{0,i}$ and increase the Rabi frequency $\Omega$ to $4E_R/\hbar$ according to $\hbar\Omega(t)/E_R = 4\sqrt{(2-t/\tau_1)t/\tau_1}$ in a time $\tau_{1}$. In step (2), we sweep the detuning to $\delta_{0,f}$ in a time $\tau_{2}$ while simultaneously increasing $\Omega$ to its final value $\Omega_{f}$. For all experiments, we set $\hbar\Omega_f/E_R$ in the range $4.5-5.3$. At these moderate values of the Rabi frequency the dispersion relation has a single minimum and losses due to inelastic photon scattering from the Raman beams remain manageable. For $\hbar\Omega/E_R=4.5$ and $\hbar\delta_0/E_R=-28.9$, we measure a single-body lifetime for the Raman-dressed BEC of $\tau=38(2)$ ms ($1/\mathrm{e}$ value), in agreement with \emph{ab initio} calculations \cite{Wei2013}.

\begin{figure}[t!]
\centering
\includegraphics[trim={6.1cm 11.7015cm 6.1cm 11.7015cm},clip]{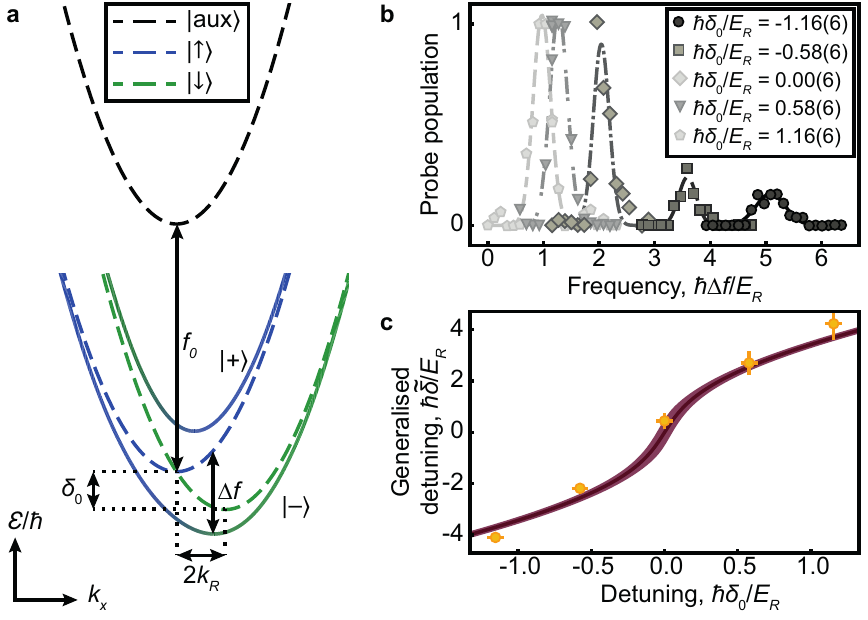}
\caption{{\bf Calibration of mechanical momentum by spin ejection spectroscopy. } \textbf{a}, Dispersion relation of the dressed states $\ket{-}$ and $\ket{+}$, and of the bare states $\ket{\uparrow},\ket{\downarrow}$, and $\ket{\text{aux}}$ in the rotating frame. Due to the Raman coupling, states $\ket{\uparrow}$ and $\ket{\downarrow}$ are shifted by $2k_R$ in momentum and $\delta_0$ in frequency.
We exploit the $\ket{-}$ to $\ket{\text{aux}}$ transition, of frequency $f_0+\Delta f$, to extract the mechanical momentum $k_{\text{mech}}$ of the cloud. \textbf{b}, rf spectroscopy of the $\ket{-}$ to $\ket{\text{aux}}$ transition for various values of the two-photon Raman detuning $\delta_0$ with $\hbar\Omega/E_R=4.5(3)$, corresponding to the parameters of the expansion experiments of Fig. \ref{Fig:Fig4}\textbf{b}. \textbf{c}, Value of the generalised detuning $\hbar\tilde{\delta}/E_R=\hbar\delta_0/E_R-4k_x/k_R$ vs. $\delta_0$ (circles) extracted from the measured frequency difference $\Delta f=(\tilde{\Omega}-\tilde{\delta})/2$. Solid line: theory prediction for $k_{\text{mech}}/k_R=0$. Shaded area: uncertainty in the theoretical value caused by the uncertainties in $\delta_0$ and $\Omega$. We conclude that the preparation procedure for the experiments of Fig. \ref{Fig:Fig4} imparts negligible mechanical momentum to the cloud.}
\label{Fig:EFig1}
\end{figure}

We perform an independent set of experiments to verify that this procedure prepares the atoms in the minimum of the dispersion relation $k_{\text{min}}$, imparting to them negligible mechanical momentum $k_{\text{mech}}$. This crosscheck is important because a finite value of $k_{\text{mech}}$ can modify the interpretation of the data due to the momentum dependence of $\hbar\tilde{\delta}/E_R=\hbar\delta_0/E_R-4k_x/k_R$, where $k_x=k_{\text{min}}+k_{\text{mech}}$. To this end, after preparing the system into state $\ket{-}$, we apply an rf pulse to transfer the atoms to a previously unoccupied bare state $\ket{\text{aux}}\equiv\ket{F=1,m_F=-1}$. This spin ejection spectroscopy scheme \cite{Cheuk2012, Wang2012} yields $\tilde{\delta}$ from the rf frequency with the highest ejection probability, which as depicted in Fig. \ref{Fig:EFig1}\textbf{a} happens at $f_0+\Delta f=f_{0}+(\tilde{\Omega}-\tilde{\delta})/2$, where $f_0$ is the frequency of the $\ket{\uparrow}$ to $\ket{\text{aux}}$ transition. For the parameters of Fig. \ref{Fig:Fig4}\textbf{b}, we obtain $\hbar\tilde{\delta}/E_R=[-2.18(17),0.45(30),2.70(48)]$ for $\delta_0=[-0.58(6),0.00(6),+0.58(6)]$, see Fig.~\ref{Fig:EFig1}\textbf{c}. Therefore, the inferred mechanical momenta $k_{\text{mech}}/k_R=[-0.09(8),-0.11(20),-0.04(16)]$ are compatible with zero for all of our expansion measurements. For the experiments of Fig. \ref{Fig:Fig2} and \ref{Fig:Fig3}\textbf{a}, an analogous measurement yields a mechanical momentum $k_{\text{mech}}/k_R=0.05(34)$, which agrees with the value determined from the velocity of the atomic cloud in the absence of the Bragg kick $k_{\text{mech}}/k_R=0.16(25)$ and is also compatible with zero.
\\

{\setlength{\parindent}{0pt}\textbf{Momentum kick.}} For the experiments of Figs. \ref{Fig:Fig2} and \ref{Fig:Fig3}, we impart momentum to the BEC \emph{via} Bragg diffraction. To this end, we briefly pulse on the atoms two counter-propagating laser beams of wavelength $\lambda_{B}=1064$ nm and adjustable frequency difference.  Depending on the sign of the latter, $88\,\%$ of the atoms are transferred to the momentum class $+2 k_B$ or $-2k_B$ along the $x$ axis by a Bragg pulse, where $k_B=2\pi/\lambda_{B}$.
\\

{\setlength{\parindent}{0pt}\textbf{Soliton experiments.}}
For the soliton experiments of Fig. \ref{Fig:Fig3}, we create a potential barrier with a blue-detuned elliptical laser beam of wavelength $\lambda=765$ nm and $1/\text{e}^{2}$ radii $(w_{x},w_{y}) = (14, 350)\,\mu$m in the atomic plane. The beam propagates along the $z$ axis and its power is set such that the barrier height exceeds the kinetic energy of the atoms (in practice, such that a BEC is reflected from it).

In Fig. \ref{Fig:Fig3}\textbf{b}, we produce a standard soliton starting with a BEC in state $\ket{\uparrow}$ at a magnetic field at $B_0=397$ G, where $a_{\uparrow\uparrow}/a_0=1.3$. As we remove the confining beam, we linearly decrease the applied magnetic field to $B_0=385.6$ G in $5$ ms to enter the regime of attractive interactions ($a_{\uparrow\uparrow}/a_0=-2.3$). For both chiral and standard solitons, we set an initial atom number $N$ sufficiently low to avoid their collapse \cite{Carr2002}.
\\

\subsubsection*{Data analysis}
For all experiments, we image the atomic cloud \emph{in situ} using a polarisation phase contrast scheme \cite{Cabrera2018} which gives the combined atomic density distribution of both states integrated along the $z$ axis. For the experiments of Figs. \ref{Fig:Fig2} and \ref{Fig:Fig3}, we extract the centre of mass positions $x_0,y_0$ and the widths $\sigma_x,\sigma_y$ of the cloud by fitting the images with a 2D Gaussian. For the experiments of Fig. \ref{Fig:Fig4}, we integrate the density profiles along the $y$ axis and low-pass filter the data before computing the moments of the normalised distribution $\mu_{m}=\int \text{d} x(x-\mu_{1})^{m}n(x)$, where $\mu_{1}=\int\text{d}x x n(x)$ is the centre of mass position of the atomic cloud and $\sigma=\sqrt{\mu_{2}}$ is its width. From the second and third moments, we compute the skewness parameter $s=\mu_{3}/\sigma^{3}$ that we use to characterise the asymmetry of the distribution. This observable is resilient to atom losses \cite{Chisholm2022}. Our filtering procedure is performed in Fourier space: first, we remove the DC-offset and afterwards we apply a low-pass filter with a cut-off frequency defined by the smallest value of $|k|$ which corresponds to a local minimum in the power spectral density and has a value less than $1\,\%$ of the maximum. We have verified the robustness of our data analysis procedure applying it to `false images', i.e. data from the numerical simulations to which we have added noise from the experimental images.

\subsubsection*{Validity of the chiral BF theory description}
The experiments of Fig. \ref{Fig:Fig2} and \ref{Fig:Fig3}\textbf{a} give access to parameter regimes where the encoded chiral BF Hamiltonian provides an appropriate effective description of our Raman-dressed system, and also realise situations where additional kinetic terms must be included to describe the behaviour of the system. To distinguish between these two regimes, we set $k_0/k_R=\hbar\delta_0/(4E_R)$ ($\tilde{\delta}=0$, $\tilde{\Omega}(k_0)=\Omega$) and compute the value of the expansion parameter $(q/k_R)/(\hbar\Omega/E_R)$ for the centre of mass momentum of the atomic cloud $k_{\text{COM}}$. For the $k_x>0$ data of Fig. \ref{Fig:Fig2}, the expansion parameter is $(q/k_R)/(\hbar\Omega/E_R)=2.87(44)$ and kinetic terms beyond equation \eqref{eq:Heff-full} cannot be neglected. We estimate their importance by computing the next ($4^{\text{th}}$) order correction to the kinetic energy normalised by the orders that are included in our effective model ($1^{\text{st}}$, $2^{\text{nd}}$, and $3^{\text{rd}}$), see equation \eqref{eq:Hkin}. Assuming a Gaussian density distribution for the atomic cloud, we obtain a kinetic correction $\Delta\mathcal{E}_{\text{kin}}\gtrsim4$. In contrast, for the $k_x<0$ data we obtain $(q/k_R)/(\hbar\Omega/E_R)=-0.09(35)$ and estimate $\Delta\mathcal{E}_{\text{kin}}\lesssim10^{-3}$. We expect the next order correction to the interaction energy to give a similar contribution. Therefore, we conclude that the effective chiral BF theory description is valid for the Raman-dressed solitons observed in the experiment.

\subsection*{Acknowledgements} We thank M. Ballu and J. Sanz for preparatory work on the Raman laser system, and M. Dalmonte, G. Juzeli\={u}nas, P. \"{O}hberg, L. Santos, G. Valent\'{i}-Rojas, and the ICFO Quantum Gases Experiment and Quantum Optics Theory groups for discussions. We acknowledge funding from the European Union (ERC CoG-101003295 SuperComp), MCIN/AEI/10.13039/501100011033 (LIGAS projects PID2020-112687GB-C21 at ICFO and PID2020-112687GB-C22 at UAB, and Severo Ochoa CEX2019-000910-S), Deutsche Forschungsgemeinschaft (Research Unit FOR2414, Project No. 277974659), Fundaci\'{o}n Ram\'{o}n Areces (project CODEC), Fundaci\'{o} Cellex, Fundaci\'{o} Mir-Puig, and Generalitat de Catalunya (ERDF Operational Program of Catalunya, Project QUASICAT/QuantumCat Ref. No. 001-P-001644, AGAUR and CERCA program). A. F. acknowledges support from La Caixa Foundation (ID 100010434, PhD fellowship LCF/BQ/DI18/11660040) and the European Union (Marie Sk{\l}odowska-Curie–713673), C. S. C. from the European Union (Marie Sk{\l}odowska-Curie–713729), E. N. from the European Union (Marie Sk{\l}odowska-Curie–101029996 ToPIKS), C. R. C. from an ICFO-MPQ Cellex postdoctoral fellowship, R. R. from the European Union (Marie Sk{\l}odowska-Curie–101030630 UltraComp), A. C. from the UAB Talent Research program, and L. T. from MCIN/AEI/10.13039/501100011033 and ESF (RYC-2015-17890).

\subsection*{Author Contributions}
A.F., C.S.C., E.N. and R.R. took and analysed the data. A.F., C.S.C., E.N., C.R.C. and R.R. prepared the experiment. A.C. and L.T. developed the theory. C.S.C. performed the numerical simulations. L.T. supervised the work. All authors contributed extensively to the interpretation of the results and to the preparation of the manuscript.

\subsection*{Author Information}
The authors declare no competing financial interests. Correspondence and requests for materials should be addressed to L.T.  (leticia.tarruell@icfo.eu) and A.C. (alessio.celi@uab.cat) for experiment and theory, respectively.

\end{document}